\documentclass[a4paper,11pt]{scrartcl}
\pdfoutput=1
\usepackage{amsmath,amssymb}
\usepackage{hyperref}
\usepackage{color}
\usepackage{bbm}
\usepackage{graphicx}
\allowdisplaybreaks[3]
\usepackage{multirow}

\newcommand{\GeV}{\,{\rm GeV}}

\newcommand{\tsum}{{\textstyle\sum}}
\newcommand{\be}{\begin{equation}}
\newcommand{\ee}{\end{equation}}
\newcommand{\bea}{\begin{eqnarray}}
\newcommand{\eea}{\end{eqnarray}}

\addtokomafont{disposition}{\rmfamily\boldmath}
\let\tmptitle\title\renewcommand{\title}[1]{\tmptitle{\LARGE #1}}
\let\tmpauthor\author\renewcommand{\author}[1]{\tmpauthor{\large #1}}
\let\tmpdate\date\renewcommand{\date}[1]{\tmpdate{\normalsize #1}}
\newcommand{\abstrct}[1]{\begin{abstract}\vspace{-2em}\small\noindent#1\end{abstract}}

\title{\LARGE $B$-decay CP-asymmetries in SUSY\\ with a
$U(2)^3$ flavour symmetry}
\date{}
\author{
\large Riccardo Barbieri$^a$, Paolo Campli$^a$, Gino Isidori$^b$,
\\
\large Filippo Sala$^a$ and David M. Straub$^a$
\\[0.5 cm]
$^a${\normalsize\itshape Scuola Normale Superiore and INFN, Piazza dei Cavalieri 7, 56126 Pisa, Italy}\\
$^b${\normalsize\itshape INFN, Laboratori Nazionali di Frascati, Via E.~Fermi 40, 00044 Frascati, Italy}
}

\begin{document}

\maketitle

\abstrct{\noindent
We study CP asymmetries in rare $B$ decays within supersymmetry with a $U(2)^3$ flavour symmetry, motivated by the SUSY flavour and CP problems, the hierarchies in the Yukawa couplings and the absence so far of any direct evidence for SUSY.
Even in the absence of flavour-blind phases, we find potentially sizable CP violating contributions to $b\to s$ decay amplitudes.
The effects in the mixing-induced CP asymmetries in $B\to\phi K_S$ and $B\to\eta' K_S$, angular CP asymmetries in $B\to K^*\mu^+\mu^-$ and the direct CP asymmetry in $B\to X_s\gamma$ can be in the region to be probed by LHCb and next generation $B$ factories.
At the same time, these effects in B decays are compatible with CP violating contributions to meson mixing, including a non-standard $B_s$ mixing phase hinted by current tensions in the CKM fit mostly between $S_{\psi K_S}, \epsilon_K$ and $\Delta M_{B_s}/\Delta M_{B_d}$.
}

\section{Introduction}

Weak scale supersymmetry has long been the leading candidate for physics beyond the Standard Model (SM), but it is increasingly challenged by the lack of any clear experimental evidence in its favour. Arguably, the most important challenges to be explained are the success of the CKM description of flavour violation, the tight bounds on flavour-blind CP violation from electric dipole moment (EDM) searches and the tightening sparticle mass bounds from LHC.

A promising approach to address these three problems is to make use of the fact that the third generation of quarks is special, since it has sizable Yukawa couplings and is weakly mixed with the first two generations. If the first two generations of quarks and squarks form doublets under a $U(2)$ flavour symmetry commuting with the gauge group, the flavour problem is ameliorated while at the same time the hierarchies in the Yukawa couplings can be (at least partially) understood \cite{Pomarol:1995xc,Barbieri:1995uv}. Furthermore, a flavour symmetry for the light quark generations fits well to a hierarchy in the spectrum of squarks, with the third generation partners being light and the first two generation ones in the multi-TeV region \cite{Cohen:1996vb,Dimopoulos:1995mi,Giudice:2008uk,Barbieri:2010ar}. This has two virtues, in addition to further ameliorating the flavour problem: First, it solves the CP problem, decoupling the contributions of flavour-blind phases to the observable EDMs, which involve first generation fermions \cite{Barbieri:2011vn}. Second, it evades the strongest sparticle direct search bounds. While the lower bounds on first two generation squark masses from LHC are approaching a TeV, third generation squarks can still be in the few-hundred GeV region for gluino masses above about $600$~GeV, depending on the gluino decay branching ratios and on the lightest neutralino mass.\footnote{For a first LHC analysis with about $1~\text{fb}^{-1}$ of integrated luminosity, see \cite{Aad:2011ks,ATLAS-CONF-2011-098}.}

Motivated by these facts, a $U(2)^3$ symmetry has been considered in \cite{Barbieri:2011ci} together with a hierarchical squark spectrum and has been shown to solve, in addition to the above-mentioned problems, tensions present in the fits of the CKM matrix by contributing to CP violation in meson mixing. In this way, a non-standard $B_s$ mixing phase was predicted, as currently hinted by Tevatron data \cite{Abazov:2010hv,Lenz:2010gu}.

The purpose of this article is to extend the analysis in \cite{Barbieri:2011ci} to $\Delta F=1$ 
processes, i.e.~rare decays, studying in particular possible signatures of CP violation 
correlated with the predicted CP violation in meson mixing ($\Delta F=2$).
Contrary to the  $\Delta F=2$ sector, where the pattern of deviations from the SM 
identified in~\cite{Barbieri:2011ci} is unambiguously dictated by the $U(2)^3$ symmetry, the predictions of $\Delta F=1$ observables are more model dependent. 
In this analysis we concentrate on a framework with moderate values of $\tan\beta$
and, in order to establish a link between 
$\Delta F=2$ and $\Delta F=1$ CP-violating observables, 
we take small flavour-blind phases and assume
the dominance of gluino-mediated flavour-changing amplitudes.

\section{Setup}

We briefly review the $U(2)^3$ framework, referring to \cite{Barbieri:2011ci} for details and derivations. We consider a global flavour symmetry
\begin{equation}
G_F=U(2)_Q\times U(2)_u\times U(2)_d
\end{equation}
broken {\em minimally} by three spurions, transforming as $\Delta Y_u= (2, \bar{2}, 1)$,
$\Delta Y_d= (2, 1, \bar{2})$ and $V = (2,1,1)$, respectively. This breaking pattern leads to non-minimal flavour violation in the down quark-squark-gluino vertex,
\begin{equation}
(\bar{d}_{L,R} W^d_{L,R} \tilde{d}_{L,R}) \tilde{g}.
\end{equation}
which can be approximately written as
\begin{equation}
 W^d_L
= \left(\begin{array}{ccc}
 c_d &  \kappa^* & - \kappa^*  s_L e^{i\gamma_L}  \\
- \kappa  &  c_d & -c_d s_L e^{i\gamma_L}   \\
  0  &  s_L e^{-i\gamma_L} & 1 \\
\end{array}\right),
\qquad
W^d_R = \mathbbm 1 \,,
\label{eq:WL}
\end{equation}
where  $\kappa =  c_d V_{td}/V_{ts}$ and $c_d$ is fixed by the CKM matrix, which in these conventions reads
\begin{equation}
 V_\text{CKM}=\left(\begin{array}{ccc}
 1- \lambda^2/2 &  \lambda & s_u s e^{-i \delta}  \\
-\lambda & 1- \lambda^2/2   & c_u s  \\
-s_d s \,e^{i (\phi+\delta)} & -s c_d & 1 \\
\end{array}\right),
\label{eq:CKM}
\end{equation}
where
$s_uc_d - c_u s_d e^{-i\phi}  = \lambda e^{i \delta}$. A direct fit of this parametrization to tree-level observables results in
\begin{align}
s_u &= 0.086\pm0.003
\,,&
s_d &= -0.22\pm0.01
\,,\\
s &= 0.0411 \pm 0.0005
\,,&
\phi &= (-97\pm 9)^\circ
\,.
\end{align}
The mixing $s_L$ and the phase\footnote{%
Our phase $\gamma_L$ is equivalent to the phase $\gamma$ in \cite{Barbieri:2011ci}.
} $\gamma_L$ in (\ref{eq:WL}), on the other hand, are free parameters. Defining
\begin{equation}
\xi_L = \dfrac{c_d s_L}{|V_{ts}|}e^{i\gamma_L} \,,
\end{equation}
the mixing amplitudes in the $K$, $B_d$ and $B_s$ meson systems are modified as follows,
\begin{align}
M_{12}^K &= \left(M_{12}^K\right)_\text{SM}^{(tt)} \left(1+|\xi_L|^4 F_0\right) + \left(M_{12}^K\right)_\text{SM}^{(tc+cc)}
,\\
M_{12}^{B_{d,s}} &= \left(M_{12}^{B_{d,s}}\right)_\text{SM} \left(1+\xi_L^2 F_0\right)
,
\end{align}
where $F_0>0$ is a loop function, predicting a universal shift in the $B_d$ and $B_s$ mixing phases.
The mixing induced CP asymmetries in $B\to\psi K_S$ and $B_s\to\psi\phi$ are thus
\begin{equation}
S_{\psi K_S} =\sin\left(2\beta + \phi_\Delta\right)
\,,\qquad
S_{\psi\phi} =\sin\left(2|\beta_s| - \phi_\Delta\right)
\,,
\label{eq:phid}
\end{equation}
where $\phi_\Delta=\text{arg}\!\left(1+\xi_L^2 F_0\right)$.

The current tensions in the CKM fit arising mostly between $S_{\psi K_S}$, $|\epsilon_K|$ and $\Delta M_s/\Delta M_d$ can be accommodated by the shifts in $S_{\psi K_S}$ and $|\epsilon_K|$. Assuming this to be the case,
a global fit of the CKM matrix to $\Delta F=2$ observables including the supersymmetric contributions leads to the following predictions at the $90\%$~C.L.,
\begin{gather}
0.8 < |\xi_L| < 2.1
\label{eq:xiL}
\,,\\
-9^\circ < \phi_\Delta < -1^\circ
\,,\\
-86^\circ < \gamma_L < -25^\circ
\text{ ~or~ }
94^\circ < \gamma_L < 155^\circ
\label{eq:gamma}
\,,\\
0.05 < S_{\psi\phi} < 0.20,
\,.
\end{gather}
with  gluino and  left-handed sbottom masses necessarily below $1\div 1.5$ TeV.
The ambiguity in $\gamma_L$ stems from the fact that the $\Delta F=2$ observables are only sensitive to the phase of $\xi_L^2$, and thus to $2\gamma_L$.

The presence of the phase $\gamma_L$ leads to contributions to CP asymmetries in $B$ physics. Additional contributions can arise from flavour-blind phases, such as the phase of the $\mu$ term or the trilinear couplings. With hierarchical sfermions, such phases are not required to be small to meet the EDM constraints \cite{Barbieri:2011vn}. Still, to concentrate on the genuine $U(2)^3$ effects, we will take flavour blind phases to be absent in the following and thus focus on $B$ physics.

\section{$\Delta B=1$ effective Hamiltonian}\label{sec:Heff}

The part of the $b\to s$ effective Hamiltonian sensitive to NP in our setup reads
\begin{align}
 \mathcal{H}_\text{eff} &= \frac{4 G_F}{\sqrt2} V_{ts}^*V_{tb} \sum_{i=3}^{10} C_i O_i + \text{ h.c.}\,,
\end{align}
\begin{align*} 
  O_3 & =(\bar{s}P_Lb)\tsum_q(\bar{q}P_Lq) \qquad &
  O_4 & =(\bar{s}_{\alpha}P_Lb_{\beta})\tsum_q(\bar{q_{\beta}}P_Lq_{\alpha}) \\
  O_5 & =(\bar{s}P_Lb)\tsum_q(\bar{q}P_Rq) \qquad &
  O_6 & =(\bar{s}_{\alpha}P_Lb_{\beta})\tsum_q(\bar{q_{\beta}}P_Rq_{\alpha}) \\
  O_{7} & =\frac{e}{16\pi^2}m_b(\bar{s}\sigma_{\mu\nu}P_R b)F_{\mu\nu} \qquad  &
  O_{8} & =\frac{g_s}{16\pi^2}m_b(\bar{s}\sigma_{\mu\nu}P_R b)G_{\mu\nu} \\
  O_{9} & =\frac{e^2}{16\pi^2}(\bar{s}\gamma_{\mu}P_L b)(\bar{l}\gamma_{\mu}l) \qquad &
  O_{10} & =\frac{e^2}{16\pi^2}(\bar{s}\gamma_{\mu}P_L b)(\bar{l}\gamma_{\mu}\gamma_5 l) 
\end{align*}

The QCD penguin operators $O_{3\ldots6}$ are relevant for the CP asymmetries in $b\to s\bar ss$ penguin decays to be discussed below. In the case of hierarchical sfermions, the box contributions are mass-suppressed and we only have to consider gluon penguins, which contribute in a universal way as
\begin{equation}
C_3 = C_5 = -\tfrac{1}{3} C_4 = -\tfrac{1}{3} C_6 \equiv  C_G \,.
\end{equation}
The analogous photon penguin can be neglected in non-leptonic decays since it is suppressed by $\alpha_{em}/\alpha_s$, but it can contribute to $C_9$. There is no effect in $C_{10}$, which remains SM-like.

In the MSSM, the $\Delta F=1$ effective Hamiltonian receives contributions from loops involving charginos, neutralinos, charged Higgs bosons or gluinos. In the following, we will concentrate for simplicity on gluino contributions, which are always proportional to the complex $\xi_L$ in $U(2)^3$. Among the remaining contributions, some are proportional to $\xi_L$ while some are real in the absence of flavour blind phases. Their omission does not qualitatively change our predictions for CP asymmetries,  but we stress that the real contributions can have an impact in particular on the branching ratios to be considered below.

The gluino contributions to $C_G$ and $C_9$ are
\begin{equation}
C_G  = - \xi_L \; \dfrac{\alpha_s}{\alpha_2} \; \dfrac{\alpha_s}{4 \pi} \; \dfrac{m_W^2}{m_{\tilde{b}}^2} \;
\dfrac{13}{108} f_G(x_{\tilde{g}})
\,, \qquad
C_9 = \xi_L \; \dfrac{\alpha_s}{\alpha_2} \; \dfrac{m_W^2}{m_{\tilde{b}}^2} \;
\dfrac{2}{27} f_{\gamma}(x_{\tilde{g}})
\,,
\end{equation}
where here and throughout, $x_{\tilde{g}} = m_{\tilde{g}}^2/m_{\tilde{b}}^2$ and all the loop functions are defined such that $f_i(1)=1$ with the exact form given in appendix~\ref{sec:LF}. 

The main difference concerning the magnetic and chromomagnetic Wilson coefficients $C_{7}$ and $C_{8}$ is that here $\tilde{b}_L$-$\tilde{b}_R$
mass insertion contributions are important, while for the other Wilson coefficients they were
chirality-suppressed. The gluino contributions read
\begin{align}
 C_{7} & =  - \xi_L \; \dfrac{\alpha_s}{\alpha_2} \; \dfrac{m_W^2}{m_{\tilde{b}}^2}\;
 \dfrac{1}{27} \left[ f_{7}(x_{\tilde{g}}) + 2 \dfrac{\mu \tan \beta - A_b}{m_{\tilde{g}}}
 g_{7}(x_{\tilde{g}}) \right]
 \,,\\
 C_{8} & =  - \xi_L \; \dfrac{\alpha_s}{\alpha_2} \; \dfrac{m_W^2}{m_{\tilde{b}}^2} \;
 \dfrac{5}{36} \left[ f_{8}(x_{\tilde{g}}) + 2 \dfrac{\mu \tan \beta - A_b}{m_{\tilde{g}}}
 g_{8}(x_{\tilde{g}}) \right]
 \,.
\end{align}

For $m_{\tilde g}=m_{\tilde b}\equiv\tilde m$, we thus find at the scale $m_W$
\begin{align}
 C_G & =  -1.1\times10^{-4}
 \left( \dfrac{500 \GeV}{\tilde m}\right)^2 \xi_L
\,,\\
 C_{\gamma} & =  9\times10^{-3} \left( \dfrac{500 \GeV}{\tilde m}\right)^2 \xi_L
\,,\\
 C_{7} & =  -3.4\times10^{-3} \left( \dfrac{500 \GeV}{\tilde m}\right)^2 \xi_L
 \left[1 + 2 \dfrac{\mu \tan \beta - A_b}{\tilde m} \right]
,\\
 C_{8} & =  -1.3\times10^{-2} \left( \dfrac{500 \GeV}{\tilde m}\right)^2 \xi_L
 \left[ 1 + 2 \dfrac{\mu \tan \beta - A_b}{\tilde m}  \right]
.
\end{align}

A model-independent consequence of the $U(2)^3$ symmetry
is that the modification of $b\to s$ and $b\to d$ $\Delta F=1$ amplitudes is {\em universal}, i.e.~only distinguished by the same CKM factors as in the SM (exactly as in the  $U(3)^3$, 
or MFV case~\cite{D'Ambrosio:2002ex}). Therefore, all the expressions for the $b\to s$ Wilson coefficients 
derived in this section are also valid for $b\to d$ processes.

\section{$B$ physics observables}

\begin{table}[tp]
\renewcommand{\arraystretch}{1.3}
 \begin{center}
\begin{tabular}{llll}
\hline
Observable & SM prediction & Experiment & Future sensitivity\\
\hline
$\text{BR}(B\to X_s\gamma)$ & $(3.15\pm0.23)\times10^{-4}$ \cite{Misiak:2006zs} & $(3.52\pm0.25)\times10^{-4}$ & $\pm0.15\times10^{-4}$  \\
$A_\text{CP}(b\to s\gamma)$ &  $(-0.6\div2.8)\%$ \cite{Benzke:2010tq} & $(-1.2 \pm 2.8) \%$ & $\pm0.5\%$ 
\\
$\text{BR}(B\to X_d\gamma)$ & $(1.54^{+0.26}_{-0.31})\times10^{-5}$ \cite{Crivellin:2011ba} & $(1.41\pm0.49)\times10^{-5}$ &
\\
$S_{\phi K_S}$  & $0.68\pm0.04$ \cite{Buchalla:2005us,Beneke:2005pu} & $0.56^{+0.16}_{-0.18}$ & $\pm0.02$ \\
$S_{\eta' K_S}$ & $0.66\pm0.03$ \cite{Buchalla:2005us,Beneke:2005pu} & $0.59\pm0.07$  & $\pm0.01$ \\
$\langle A_7 \rangle$ & $(3.4\pm0.5)\times10^{-3}$ \cite{Altmannshofer:2008dz} & -- &  \\
$\langle A_8 \rangle$ & $(-2.6\pm0.4)\times10^{-3}$ \cite{Altmannshofer:2008dz} & -- &  \\
\hline
 \end{tabular}
 \end{center}
\renewcommand{\arraystretch}{1}
 \caption{SM predictions, current experimental world averages \cite{Asner:2010qj} and experimental sensitivity at planned experiments \cite{Aushev:2010bq,O'Leary:2010af} for the $B$ physics observables. $<A_{7,8}>$ are suitable angular CPV asymmetries in $B\rightarrow K^* \mu^+ \mu^-$.}
 \label{tab:exp}
\end{table}

\subsection{BR($B\to X_q\gamma$)}\label{sec:bsg}

The branching ratio of $B\to X_s\gamma$ is one of the most important flavour constraints in the MSSM in view of the good agreement between theory and experiment. Experimentally, the quantities
\begin{equation}
R_{bq\gamma} = \frac{\text{BR}({B} \rightarrow X_q \gamma)}{\text{BR}({B} \rightarrow X_q \gamma)_\text{SM}}
\end{equation}
are constrained to be
\begin{equation}
R_{bs\gamma}=1.13\pm0.11
\,,\qquad
R_{bd\gamma}=0.92\pm0.40
\,,
\label{eq:Rbqg}
\end{equation}
using the numbers in table~\ref{tab:exp}. In $U(2)^3$, one has $R_{bs\gamma}\approx R_{bd\gamma}$ just as in MFV, so the $b\to s\gamma$ constraint is more important and we will concentrate on it in the following.

Beyond the SM (but in the absence of right-handed currents), the branching ratio can be written as 
\begin{equation}
\text{BR}({B} \rightarrow X_s \gamma) = R \left[ \left|C_{7}^\text{SM,eff} + C_{7}^\text{NP,eff} \right|^2
+ N(E_{\gamma}) \right],
\end{equation}
where $R = 2.47 \times 10^{-3}$ and $N(E_{\gamma}) = (3.6 \pm 0.6) \times 10^{-3}$ for a photon energy cut-off $E_{\gamma} = 1.6$ \GeV\ \cite{Buras:2011zb}.

Considering only gluino contributions and setting $m_{\tilde g}=m_{\tilde b}\equiv\tilde m$, we find
\begin{align}
R_{b s\gamma}  &= 1 \;+ \; 2.2 \times 10^{-2} 
\left( \dfrac{500 \GeV}{\tilde m}\right)^2
|\xi_L| \cos \gamma_L
\left( 1
+ 2 \dfrac{\mu \tan \beta - A_b}{\tilde m}  \right)
.
\end{align}

As stressed in section~\ref{sec:Heff}, there are additional real contributions to the Wilson coefficient $C_{7,8}$ that can modify the branching ratio. In particular, there is a $\tan\beta$ enhanced chargino contribution proportional to the stop trilinear coupling, which can interfere constructively or destructively with the SM. Thus, with a certain degree of fine-tuning, the constraints in \ref{eq:Rbqg} can always be fulfilled. In our numerical analysis, we will require the branching ratio including {\em only SM and gluino contributions} to be within $3\sigma$ of the experimental measurement.

\subsection{$A_\text{CP}(B\to X_s\gamma)$}\label{sec:Absg}

The direct CP asymmetry in $B\to X_s\gamma$
\begin{equation}
A_\text{CP}(B\to X_s\gamma)= \dfrac{\Gamma(\bar{B} \rightarrow X_s \gamma) - \Gamma(B\rightarrow X_{\bar{s}} \gamma)}
{\Gamma(\bar{B} \rightarrow X_s \gamma) + \Gamma(B \rightarrow X_{\bar{s}} \gamma)}
\,,
\end{equation}
already constrained by the $B$ factories as shown in table~\ref{tab:exp}, will be measured by next generation experiments to a precision of 0.5\%. On the theory side, the recent inclusion of ``resolved photon'' contributions reduced the attainable sensitivity to NP in view of the large non-perturbative SM contribution leading to \cite{Benzke:2010tq}
\begin{equation}
-0.6\% <A_\text{CP}(B\to X_s\gamma)_\text{SM} < 2.8\%
\end{equation}
compared to an earlier estimate of the short-distance part \cite{Hurth:2003dk}
\begin{equation}
 A_\text{CP}(B\to X_s\gamma)_\text{SM}^\text{SD} = (0.44^{+0.24}_{-0.14})\% \,.
\end{equation}

The new physics contribution can be written as
\begin{equation}
 A_\text{CP}(B\to X_s\gamma)_\text{NP} \times R_{b s\gamma} = -0.29 \text{Im}(C_{7}^{\text{NP}}) + 0.30 \text{Im}(C_{8}^\text{NP})
-0.99 \text{Im}(C_{7}^{\text{NP}*}C_{8}^\text{NP}),
\end{equation}
valid for $E_{\gamma}=1.85 \GeV$.

Setting $m_{\tilde g}=m_{\tilde b}\equiv\tilde m$ and $R_{bs\gamma}=1$, the gluino contributions to the CP asymmetry are
\begin{align}
 A_\text{CP}(B\to X_s\gamma)_\text{NP} = 
- 1.74 \times 10^{-3} \left( \dfrac{500 \GeV}{\tilde m}\right)^2
 |\xi_L| \sin \gamma_L
\left(1
+ 2 \dfrac{\mu \tan \beta - A_b}{\tilde m} \right)
\end{align}
we note that the CP asymmetry can have either sign due to the two solutions for $\gamma_L$ allowed by $\Delta F=2$, see (\ref{eq:gamma}).

\subsection{$B \rightarrow K^* \mu^+ \mu^-$}

Angular CP asymmetries in $B \rightarrow K^* \mu^+ \mu^-$ are sensitive probes of non-standard CP violation and will be measured soon at the LHCb experiment. In our framework, where right-handed currents are absent, the relevant observables are the T-odd CP asymmetries $A_7$ and $A_8$ as defined in \cite{Altmannshofer:2008dz}.

For these observables, integrated in the low dilepton invariant mass region, we obtain the simple dependence on the Wilson coefficients, as usual to be evaluated at the scale $m_b$,
\begin{align}
\langle A_7 \rangle \times R_\text{BR} & \approx  - 0.71 \; \text{Im} (C_{7}^\text{NP})
\,,\\
\langle A_8 \rangle \times R_\text{BR}  & \approx    0.40 \; \text{Im} (C_{7}^\text{NP}) + 0.03 \; \text{Im} (C_{9}^\text{NP})
\,,
\end{align}
where $R_\text{BR}$ is the ratio between the full result for the CP-averaged branching ratio and the SM one \cite{Barbieri:2011vn}, $R_\text{BR} \approx 1$ in our framework.
Although $\text{Im} (C_{7}^\text{NP})$ and $\text{Im} (C_{9}^\text{NP})$ can be of the same order, the contribution from $C_{9}^\text{NP}$ is numerically suppressed and one will thus still have approximately
\begin{equation}
\langle A_8 \rangle \simeq - 0.56 \langle A_7 \rangle
\,.
\end{equation}

Setting $m_{\tilde g}=m_{\tilde b}\equiv\tilde m$ and $R_\text{BR}=1$, the gluino contributions to $\langle A_7 \rangle$ read
\begin{align}
\langle A_7 \rangle
&
= 2.5 \times10^{-3} \left(
\dfrac{500 \GeV}{\tilde m} \right)^2 
|\xi_L| \sin \gamma_L
\left(1+
2 \dfrac{\mu \tan \beta - A_b}{\tilde m}
\right) \,.
\end{align}

\subsection{$S_{\phi K_S}$ and $S_{\eta' K_S}$}

The expression for the mixing-induced CP asymmetries in $B_d$ decays to final CP eigenstates $f$ is
\begin{equation}
S_f = \sin \left( 2 \beta + \phi_\Delta + \delta_f \right),
\end{equation}
where $\phi_\Delta$ is the new phase in $B_{d,s}$ mixing defined in (\ref{eq:phid}). For the tree-level decay $f=\psi K_S$, $\delta_f=0$, while for the penguin-induced modes $B\to\phi (\eta')K_S$, the contribution from the decay amplitude can be written as \cite{Buchalla:2005us}
\begin{equation}
\delta_f =
2 \text{arg}\bigg(1 + a^u_f e^{i \delta} + \sum_{i \geq 3} b_{i,f}^c C_i^\text{NP}\bigg)
\end{equation}
where $\delta=\gamma_\text{CKM}=\phi_3$ is the usual CKM angle and the $a^u_f$ and $b^c_{i,f}$ can be found in \cite{Buchalla:2005us}.

For the gluino contributions, setting $m_{\tilde g}=m_{\tilde b}\equiv\tilde m$, we obtain
\begin{align}
\sum_{i=3}^6 b_{i,\phi K_S}^c C_i^\text{NP}& = -0.73\times 10^{-2}  \left( \dfrac{500 \GeV}{\tilde m}\right)^2
|\xi_L| e^{i\gamma_L}
\,,\\
\sum_{i=3}^6 b_{i,\eta' K_S}^c C_i^\text{NP}& = -1.10\times 10^{-2}  \left( \dfrac{500 \GeV}{\tilde m}\right)^2
|\xi_L| e^{i\gamma_L}
\,,
\end{align}
\begin{align}
b_{8,\phi K_S}^c C_8^\text{NP}& = -1.82\times 10^{-2}  \left( \dfrac{500 \GeV}{\tilde m}\right)^2
|\xi_L| e^{i\gamma_L} \left(1+ 2\dfrac{\mu \tan \beta - A_b}{\tilde m}\right)
\label{eq:C8phi}
,\\
b_{8,\eta' K_S}^c C_8^\text{NP}& = -1.10\times 10^{-2}  \left( \dfrac{500 \GeV}{\tilde m}\right)^2
|\xi_L| e^{i\gamma_L} \left(1+ 2\dfrac{\mu \tan \beta - A_b}{\tilde m}\right)
\label{eq:C8eta}
.
\end{align}
The effects of the QCD and chromomagnetic penguins in the above expressions are comparable, with the exception of the left-right mixing piece only present for the chromomagnetic ones.

\begin{figure}[tbp]
\begin{center}
\includegraphics[width=0.48\textwidth]{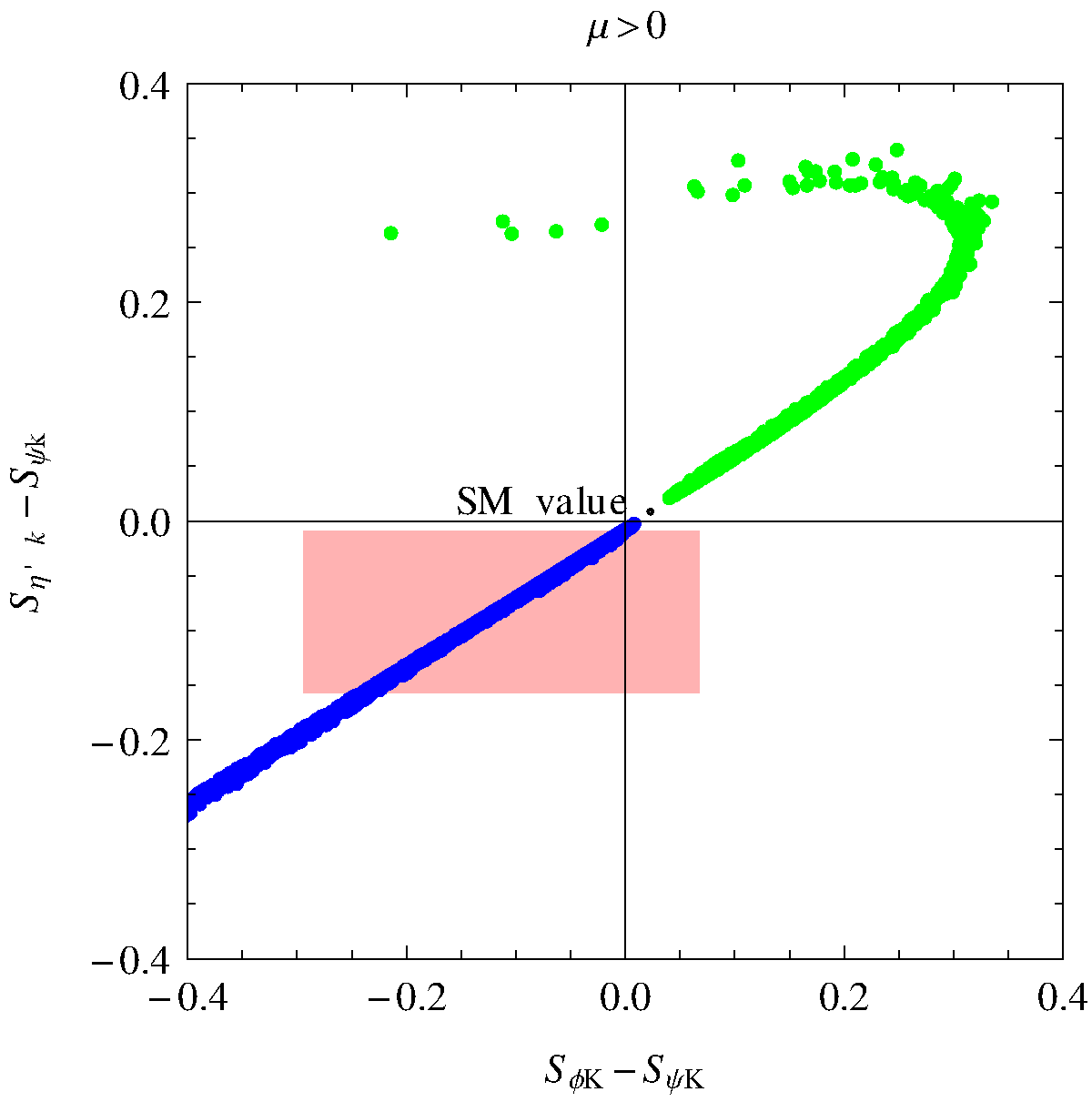}
\includegraphics[width=0.48\textwidth]{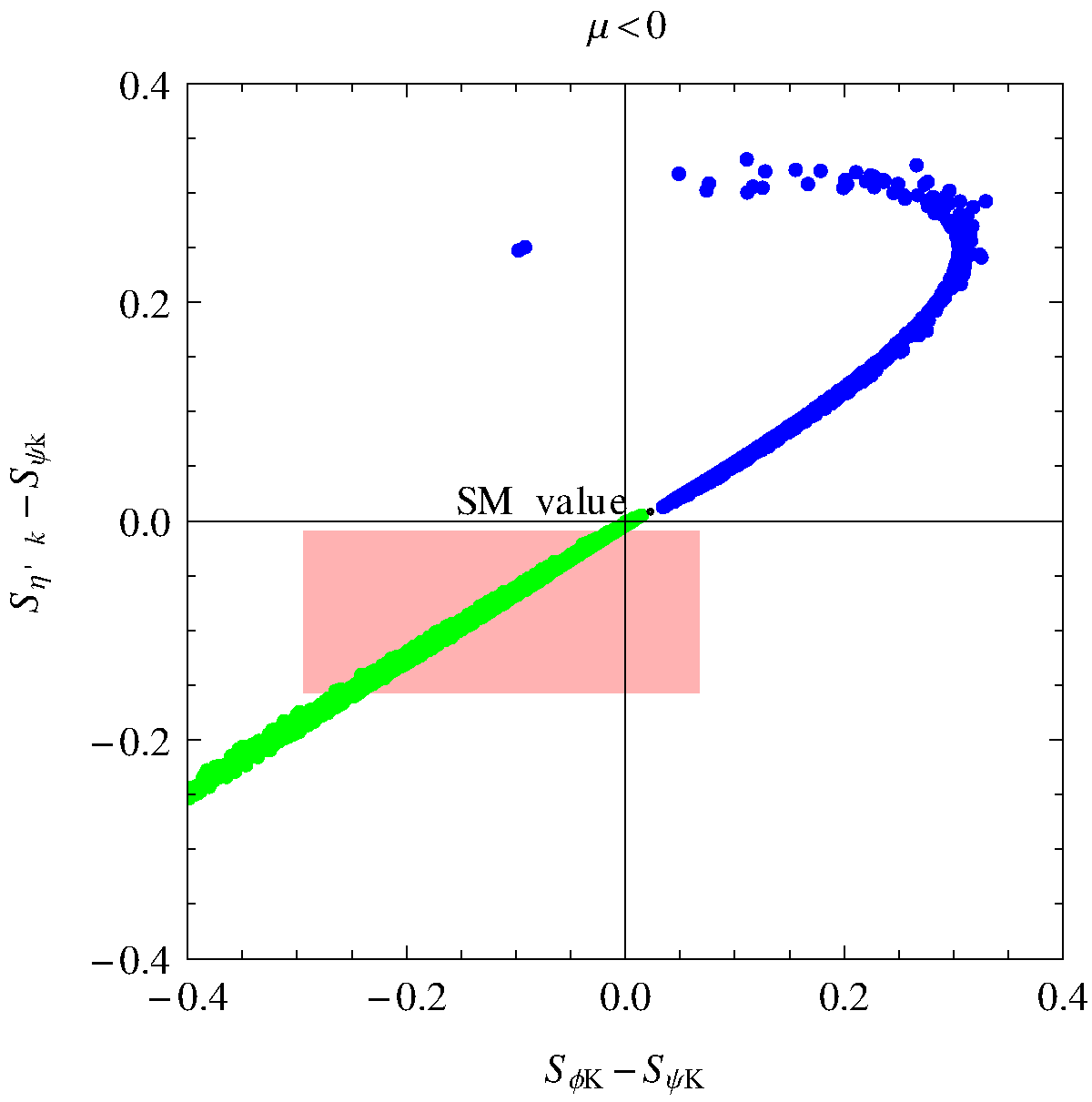}
\end{center}
\caption{Correlation between $S_{\phi K_S}-S_{\psi K_S}$ and $S_{\eta' K_S}-S_{\psi K_S}$ for positive $\mu$ (left) and negative $\mu$ (right), showing points with $\gamma_L>0$ (blue) and $\gamma_L<0$ (green). The shaded region shows the $1\sigma$ experimental ranges.}
\label{fig:S}
\end{figure} 
\begin{figure}[tbp]
\begin{center}
\includegraphics[width=0.48\textwidth]{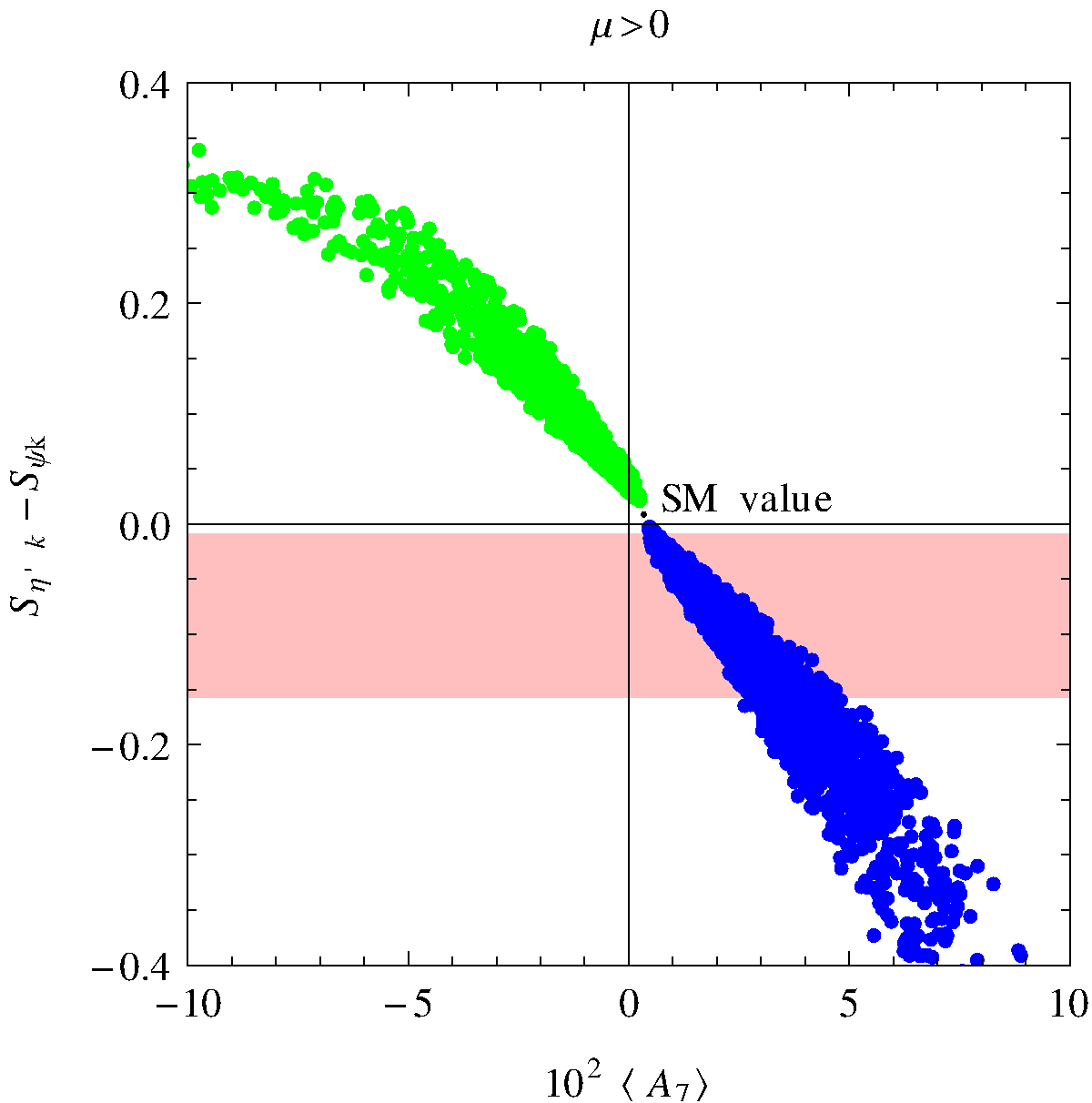}
\includegraphics[width=0.48\textwidth]{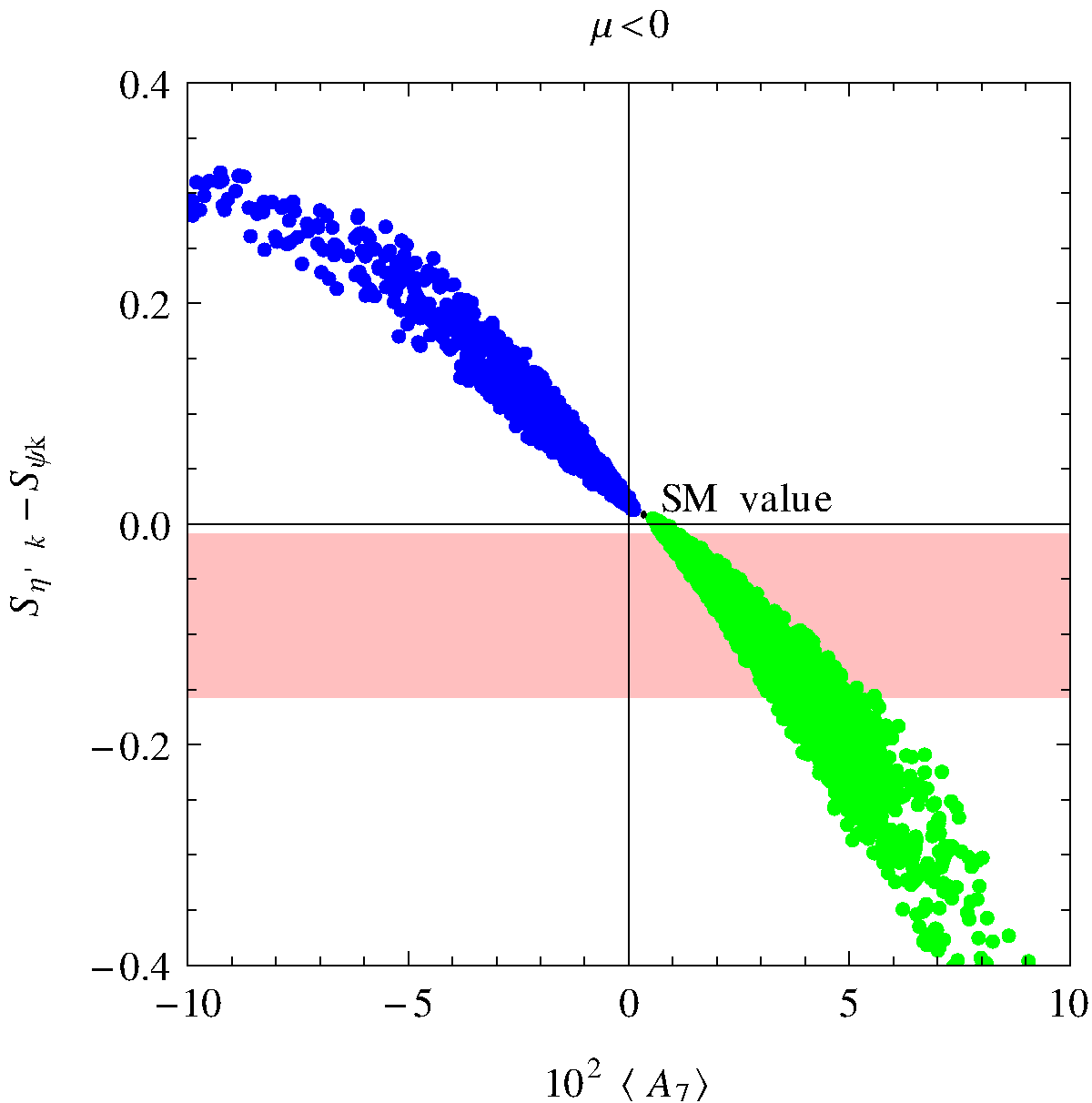}
\end{center}
\caption{Correlation between $\langle A_7 (B\to K^*\mu^+\mu^-)\rangle$ and the difference $S_{\eta' K_S}-S_{\psi K_S}$ for positive $\mu$ (left) and negative $\mu$ (right), showing points with $\gamma_L>0$ (blue) and $\gamma_L<0$ (green). The shaded region is the $1\sigma$ experimental range.
}
\label{fig:A7}
\end{figure} 
\begin{figure}[tbp]
\begin{center}
\includegraphics[width=0.48\textwidth]{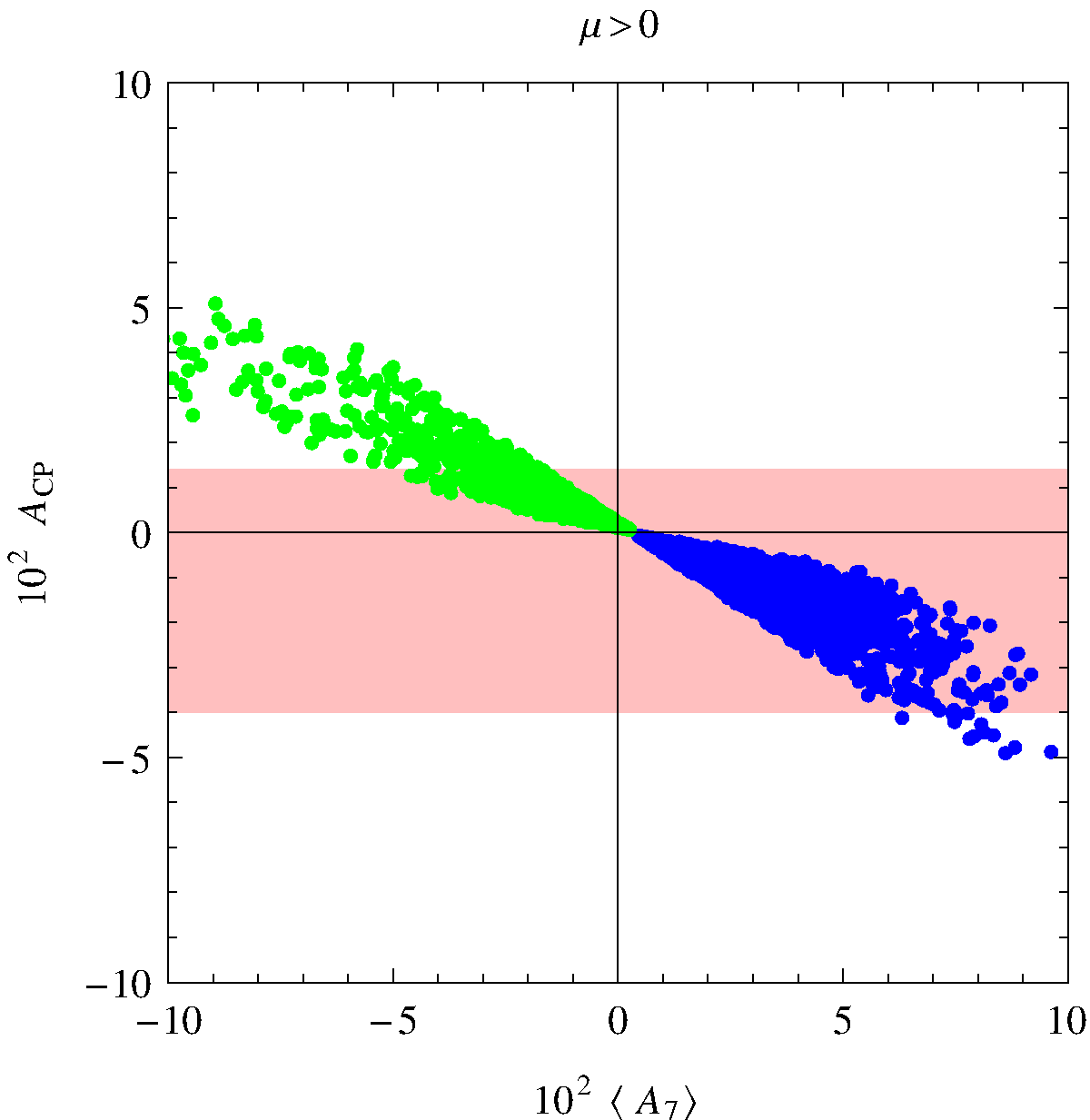}
\includegraphics[width=0.48\textwidth]{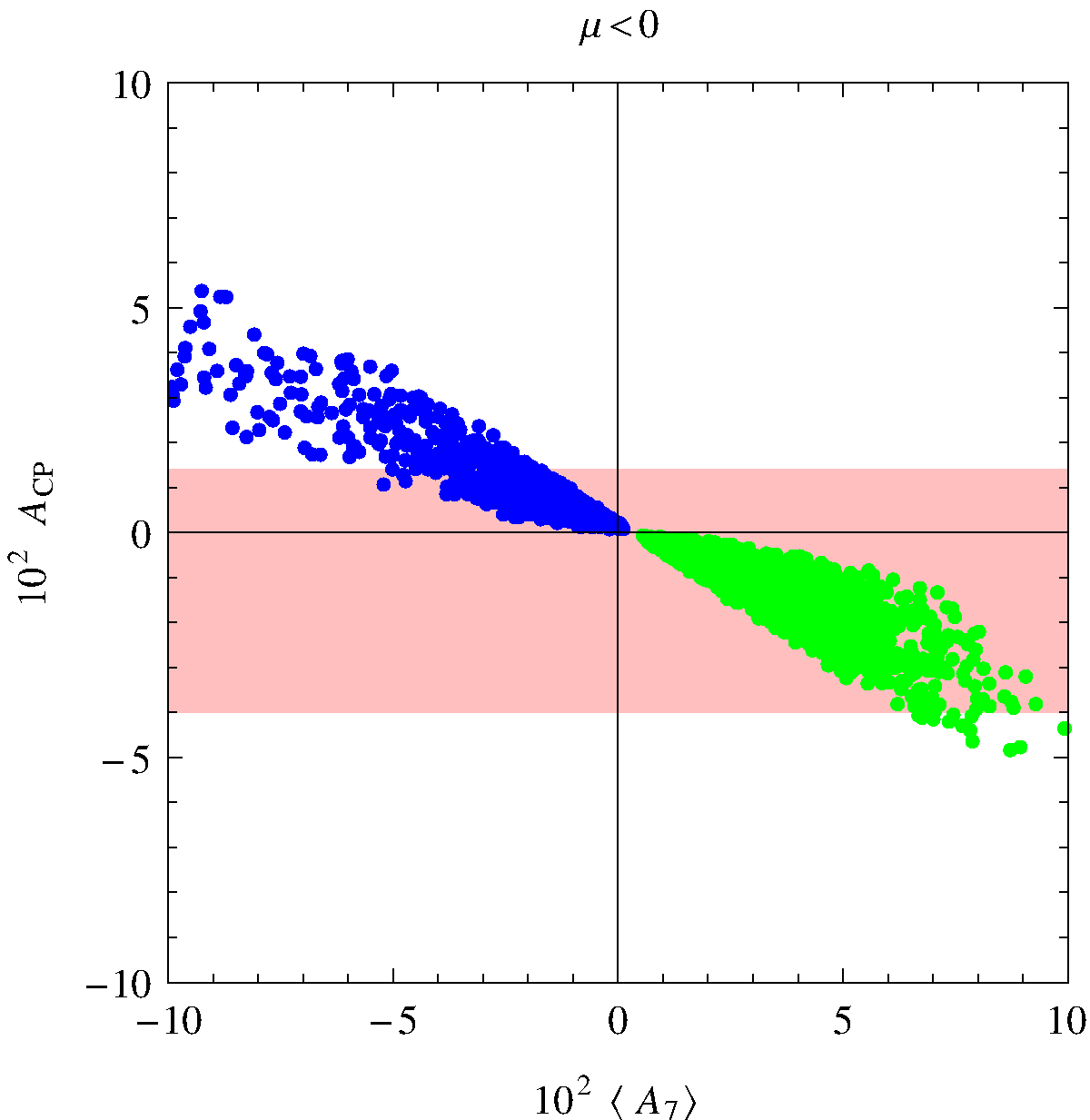}
\end{center}
\caption{Correlation between $\langle A_7 (B\to K^*\mu^+\mu^-)\rangle$ and the new physics contributions to the CP asymmetry in $B\to X_s\gamma$ for positive $\mu$ (left) and negative $\mu$ (right), showing points with $\gamma_L>0$ (blue) and $\gamma_L<0$ (green). The shaded region is the $1\sigma$ experimental range for $A_\text{CP}(B\to X_s\gamma)$, valid for vanishing SM contribution (and otherwise subject to an appropriate shift).
}
\label{fig:A7Absg}
\end{figure} 

\subsection{Numerical analysis}

In figures \ref{fig:S} and \ref{fig:A7} we show the correlations between the CP asymmetries, scanning the gluino mass between 0.5 and 1~TeV, the sbottom mass,  the $\mu$ term and $A_b$ between 0.2 and 0.5~TeV and $\tan\beta$ between 2 and 10. We require the $\Delta F=2$ observables to be in the region where the CKM tensions are reduced (cf. (\ref{eq:xiL}--\ref{eq:gamma})).
The maximum size of the effects is mostly limited by the $\text{BR}(B\to  X_s\gamma)$ constraint, which we require to be fulfilled at the $3\sigma$ level including SM and gluino contributions only (cf. the discussion at the end of sec.~\ref{sec:bsg}).

Figure~\ref{fig:S} shows the correlation between the mixing-induced CP asymmetries $S_{\phi K_S}$ and $S_{\eta' K_S}$ in relation to $S_{\psi K_S}$, effectively subtracting the contribution due to the modified $B_d$ mixing phase.
The experimental $1\sigma$ ranges corresponding to the average in table~\ref{tab:exp} are shown as shaded regions.
Due to the $\tan\beta$ enhanced terms in (\ref{eq:C8phi},~\ref{eq:C8eta}), large effects are easily possible.
A negative value for these differences, as currently indicated by the central values of the measurements, can be obtained for $\gamma_L>0$ ($\gamma_L<0$) if $\mu>0$ ($\mu<0$). For a given sign of the $\mu$ term, the sign of the $\Delta B=1$ CP asymmetries can thus serve to distinguish between the two solutions for the phase $\gamma_L$ in (\ref{eq:gamma}) allowed by the $\Delta F=2$ analysis.

Figure~\ref{fig:A7} shows the correlation between $S_{\eta' K_S}$ and the CP asymmetry $\langle A_7 \rangle$ in $B\to K^*\mu^+\mu^-$.
Values up to $\pm10\%$ would be attainable for $\langle A_7 \rangle$, while the current measurement of $S_{\eta' K_S}$ implies, at the $1\sigma$ level, $0 < \langle A_7 \rangle < 5\%$.

Finally, figure~\ref{fig:A7Absg} shows the correlation between the CP asymmetry in $B\to X_s\gamma$ and  the new physics contribution to $\langle A_7 \rangle$.
$A_\text{CP}(B\to X_s\gamma)_\text{NP}$ attains values up to $\pm5\%$. Imposing the $1\sigma$ experimental range allowed for $S_{\eta' K_S}$, this decreases to $-2\% < A_\text{CP}(B\to X_s\gamma)_\text{NP} < 0 \%$.
In the plots, we also show the $1\sigma$ experimental range for $A_\text{CP}(B\to X_s\gamma)$ (cf. table~\ref{tab:exp}), assuming a vanishing SM contribution. If the SM contribution is sizable due to long-distance effects (see section~\ref{sec:Absg}), the experimental constraints has to be shifted accordingly.

\section{Discussion and conclusions}

We have studied CP asymmetries in $B$ decays in supersymmetry with a $U(2)^3$ flavour symmetry suggested in \cite{Barbieri:2011ci} as an alternative to Minimal Flavour Violation, addressing the SUSY flavour and CP problems, partially explaining the hierarchies in the Yukawa couplings and solving tensions in the CKM fit related to CP violation in meson mixing.

Even in the absence of flavour-blind phases, we find potentially sizable CP violating contributions to $\Delta B=1$ decay amplitudes. We identify the dominant contributions to arise in the magnetic and chromomagnetic dipole operators due to their sensitivity to chirality violation, with subleading contributions in semi-leptonic and QCD penguin operators. The most promising observables are the mixing-induced CP asymmetries in non-leptonic penguin decays like $B\to\phi K_S$ or $B\to\eta' K_S$, angular CP asymmetries in $B\to K^*\mu^+\mu^-$, and the direct CP asymmetry in $B\to X_s\gamma$ 
(barring potential uncertainties in controlling the SM predictions in the radiative~\cite{Benzke:2010tq} 
and non-leptonic modes~\cite{Buchalla:2005us,Beneke:2005pu}).

Due to the different dependence on the sparticle masses, we cannot predict a clear-cut correlation between CP violating $\Delta F=1$ and $\Delta F=2$ observables. However, we have demonstrated that observable effects in $\Delta F=1$ CP asymmetries are certainly compatible with the pattern of deviations from the SM suggested by $\Delta F=2$ observables, 
if interpreted in terms of this supersymmetric framework.
Interestingly, while we considered a setup without flavour-blind phases, the correlations between $\Delta F=1$ observables turn out to be very similar to those in MFV \cite{Altmannshofer:2008hc,Altmannshofer:2009ne} or in effective MFV \cite{Barbieri:2011vn} with flavour-blind phases. The main difference between the two cases are the CP violating effects in $K$ and $B$ mixing, which occur in $U(2)^3$, but not in (effective) MFV. We view this as an interesting example of the usefulness of correlated studies of $\Delta F=1$ and $\Delta F=2$ observables as a handle to distinguish between models.
Such studies  would become extremely interesting in presence of direct evidences of 
a hierarchical sparticle spectrum from the LHC.

\section*{Acknowledgments}
This work was supported by the EU ITN ``Unification in the LHC Era'', 
contract PITN-GA-2009-237920 (UNILHC) and by MIUR under contract 2008XM9HLM.
G.I. acknowledges the support of the Technische Universit\"at M\"unchen -- Institute for Advanced
Study, funded by the German Excellence Initiative.

\appendix

\section{Loop functions}\label{sec:LF}

\begin{center}
\begin{tabular}{ll}
$f_G(x) = \dfrac{2(73 - 134 x + 37 x^2)}{39 (x-1)^3} -
\dfrac{2(18 - 27 x + x^3)}{13 (x-1)^4} \ln x$ & $f_G(1) = 1$\\
\\
$f_{\gamma}(x) = -\dfrac{2(2 - 7 x + 11 x^2)}{3(x-1)^3} +
\dfrac{4 x^3}{(x-1)^4} \ln x$ & $f_{\gamma}(1) = 1$\\
\\
$f_{7}(x) = \dfrac{2(-1 + 5 x + 2 x^2)}{(x-1)^3} -
\dfrac{12 x^2}{(x-1)^4} \ln x$ & $f_{7}(1) = 1$\\
\\
$g_{7}(x) = -\dfrac{6 x (1 + 5 x)}{(x-1)^3} +
\dfrac{12 x^2 (2 + x)}{(x-1)^4} \ln x$ & $g_{7}(1) = 1$\\
\\
$f_{8}(x) = \dfrac{-19 - 40 x + 11 x^2}{5 (x-1)^3} -
\dfrac{6 x (-9 + x)}{5 (x-1)^4} \ln x$ & $f_{8}(1) = 1$\\
\\
$g_{8}(x) = \dfrac{12 x (11 + x)}{5 (x-1)^3} +
\dfrac{6 x (-9 - 16 x + x^2)}{5 (x-1)^4} \ln x$ & $g_{8}(1) = 1$
\end{tabular}
\end{center}

\bibliographystyle{My}
\bibliography{u2df1}

\end{document}